\documentclass[3p, numbers,twocolumn,sort&compress]{elsarticle}
\usepackage{paralist}
\usepackage{amssymb,amsmath}
\usepackage{graphicx}
\usepackage{stix}
\usepackage{lineno}
\usepackage{siunitx}
\DeclareSIUnit{\ADU}{ADU}
\DeclareSIUnit{\sample}{S}
\usepackage{url}
\usepackage{subfigure}
\usepackage[colorlinks=true,urlcolor=Plum,citecolor=WildStrawberry,linkcolor=RoyalBlue]{hyperref}
\usepackage[version=4]{mhchem}
\usepackage{microtype}
\usepackage{wrapfig}
\usepackage{atlasphysics}
\usepackage{subcaption}
\usepackage{tabularx}
\usepackage{upgreek}

\newcommand{\achinos}{\textsc{ACHINOS}}
\usepackage{comment}
\begin{document}

\title{Response of a nitrogen-filled spherical proportional counter to mono-energetic neutrons}

\author[uhh]{T.~Avgitas}
\author[demo]{M.~Axiotis}
\author[uhh]{Z.~Balmforth}
\author[uhh]{P.~Gadow}
\author[saclay]{I.~Giomataris}
\author[uoa]{D.~Fassouliotis}
\author[uhh]{J.~Fock}
\author[uob]{P.~Knights}
\author[demo]{A.~Lagoyannis}
\author[uhh]{I.~Manthos}
\author[uhh]{L.~Millins}
\author[uhh,uob]{K.~Nikolopoulos}
\author[uhh]{I.~Oceano\corref{cor1}}
\ead{isabella.oceano@uni-hamburg.de}
\cortext[cor1]{Corresponding author}
\author[uoa]{C.~Prodromou}
\author[uhh]{L.~Schweder}
\author[demo]{E.~Taimpiri}
\author[uhh]{C.~Toukmenidis}
\author[demo]{A.~Ziagkova}
\author[uhh]{R.~Ward}

\address[uhh]{Institute of Experimental Physics, University of Hamburg, 22761, Hamburg, Germany}
\address[demo]{TANDEM Accelerator Laboratory, Institute of Nuclear and Particle Physics, NCSR “Demokritos”, 153 10, Athens, Greece}
\address[saclay]{IRFU, CEA, Universit\'{e} Paris-Saclay, Gif-sur-Yvette, F-91191, France}
\address[uoa]{Department of Physics, National and Kapodistrian University of Athens, 157 84, Athens, Greece}
\address[uob]{School of Physics and Astronomy, University of Birmingham, B15 2TT, United Kingdom}

\journal{Journal of Nuclear Instruments and Methods in Physics Research A}

\begin{abstract}
Neutron spectroscopy is an invaluable tool for a wide range of scientific and industrial applications, however, current approaches suffer from limitations that restrict their field of applicability. A safe and inexpensive alternative approach to neutron detection and spectroscopy is the use of a nitrogen-filled spherical proportional counter that exploits the \ce{^{14}N(n,p)^{14}C} and \ce{^{14}N(n,\alpha)^{11}B}  reactions. 
The neutron spectroscopy capabilities of the detector are demonstrated using beams of mono-energetic neutrons. A nitrogen-filled spherical proportional counter, operating at a pressure of \SI{1}{bar}, is exposed to neutrons with energies from \SIrange{0.75}{2.75}{\mega\electronvolt} at the Tandem accelerator of the National Centre for Scientific Research ``Demokritos'' in Athens. A linear energy response is observed, within the statistical precision of the measurements.
\end{abstract}


\maketitle

\section{Introduction}
\label{sec:introduction}

Neutron spectroscopy is an invaluable tool across many scientific,
industrial and medical applications~\cite{Saengkaew:2026ptq, Pietropaolo:2020frm}.

In fundamental physics, the characterisation of neutron fields is important for
rare-event searches: in underground laboratories, neutron-induced nuclear
recoils from residual cosmic-ray muons and ambient radioactivity mimic the signatures expected from dark matter
interactions~\cite{Billard:2021uyg} or neutrinoless double-$\beta$-decays~\cite{Meregaglia:2017nhx}. Such backgrounds are suppressed through passive
shielding and active veto systems~\cite{Mei:2005gm,Formaggio:2004ge}, but their precise in-situ
characterisation remains key to establish the residual contribution and
the ultimate sensitivity of the experiment. 

Beyond fundamental physics, neutron spectroscopy underpins a broad range of industrial and medical applications. 
In non-destructive analysis, energy-resolved neutron imaging maps elemental and isotopic composition through  characteristic resonances in the transmitted spectrum~\cite{10.1063/1.5136034}. 
In radiation protection and radiotherapy, the neutron energy spectrum is required to assess the biological dose, given the strongly energy-dependent neutron radiation weighting factor~\cite{ICRP103}. This applies also to  photoneutrons produced in high-energy photon-therapy beams~\cite{NASERI2010138} and to  secondary neutrons generated in proton and light-ion therapy~\cite{VEDELAGO2024107214}.
In safeguards and security, the energy spectrum of fast neutrons discriminates nuclear materials of interest from benign sources and characterises shielded assemblies~\cite{10.1063/1.3503495}.

The most widely used approach for neutron detection relies on the
\ce{^3He(n,p)^3H} reaction, which combines large thermal-neutron capture
cross-section with low
sensitivity to $\gamma$-rays. 
These properties have made \ce{^3He}
proportional counters the standard choice for thermal-neutron counting. 
When applied to fast-neutron spectroscopy, the response is degraded by the so-called wall
effect, where one or both reaction products reach the detector wall before
depositing their full energy, which results in a low-energy tail. In addition,
\ce{^3He} is a scarce resource, produced almost exclusively from the decay of
tritium: the surge in demand following the large-scale deployment of neutron
detectors for security applications has depleted the available supply and driven
up its cost, severely limiting the scalability of \ce{^3He}-based
systems~\cite{KOUZES20101035}. 

A range of alternatives has been
proposed, including \ce{BF3} and
\ce{^{10}B}-lined proportional counters, \ce{^6Li}-based scintillators, and
boron- or lithium-loaded detectors~\cite{KOUZES20101035,Simpson2011,Kouzes:2015tsc}. While effective for thermal-neutron counting,
these typically share one or more drawbacks: \ce{BF3} is toxic, coated devices have reduced efficiency and suffer from the wall effect, scintillators have increased $\gamma$-ray sensitivity and complex response
functions.
 
A promising new alternative is the use of the spherical proportional counter~\cite{Giomataris:2008ap,Knights:2025ogz}, a gaseous detector employed for light particle dark matter searches~\cite{Arnaud:2017bjh,NEWS-G:2024jms,NEWS-G:2023qwh}, as a neutron spectrometer by operating it with nitrogen~\cite{Bougamont:2015jzx}. 
To date, neutron measurements with nitrogen-filled spherical proportional counters have featured continuous energy spectra~\cite{Bougamont:2015jzx,Giomataris:2022bvz,Giomataris:2022kxw}. However, a controlled, quantitative characterisation of the detector as a neutron spectrometer requires measurements with well-defined mono-energetic neutron beams, which allows the study of the detector response  at discrete, precisely known neutron energies. 

In this work, the neutron spectroscopy capabilities of the spherical proportional counter are demonstrated using mono-energetic neutron beams. Measurements were performed at the \SI{5.5}{\mega\volt} Tandem accelerator of the National Centre for Scientific Research (NCSR) ``Demokritos'' in Athens~\cite{Harissopulos:2021lgh}. These measurements are supported by  simulations using a dedicated  framework~\cite{Katsioulas:2019sui}. 

The article is organised as follows. Section~\ref{sec:detector} introduces the nitrogen-filled spherical proportional counter and the \achinos\ multi-anode concept. Section~\ref{sec:achinos} presents the characterisation of the utilised 11-anode \achinos\ sensor. Section~\ref{sec:setup} describes the experimental configuration at NCSR ``Demokritos''. The experimental results are reported and discussed in Section~\ref{sec:results}, while Section~\ref{sec:conclusions} gives a summary.

\section{The nitrogen-filled spherical proportional counter}
\label{sec:detector}

The spherical proportional counter is a gaseous detector comprising a grounded spherical vessel that serves as the cathode, and a small spherical anode of  $\qty{1}{\milli\meter}$ radius at its centre~\cite{Giomataris:2008ap}. A high voltage is applied to the anode through a wire, which is shielded by a grounded metallic rod that also provides mechanical support. To first approximation, the magnitude of the electric field strength is radial and varies to a good approximation as $1/r^2$, where $r$ is the distance from the detector centre. This dependence naturally separates the detector volume into a large drift region, where primary ionisation electrons are guided towards the anode, and a small avalanche region near the anode surface where the field is sufficiently strong to induce charge multiplication and provide signal amplification. The detector exhibits several attractive features, including  small capacitance that is, practically, independent of the vessel radius, enabling large target masses with low electronic noise;  variable operating gas mixture and pressure; and pulse-shape discrimination capabilities that allow active volume fiducialisation and background rejection.  

When operated with a nitrogen-based gas mixture, the detector becomes sensitive to both thermal and fast neutrons through the
\begin{eqnarray*}
\ce{^{14}N + n} \to \ce{^{14}C + p} + \SI{625}{\kilo\electronvolt}\\
\ce{^{14}N + n} \to \ce{^{11}B + \alpha} - \SI{159}{\kilo\electronvolt}
\end{eqnarray*}
reactions.
\begin{figure}
  \centering
  \vspace{-0.50cm}
\includegraphics[width=0.80\linewidth]{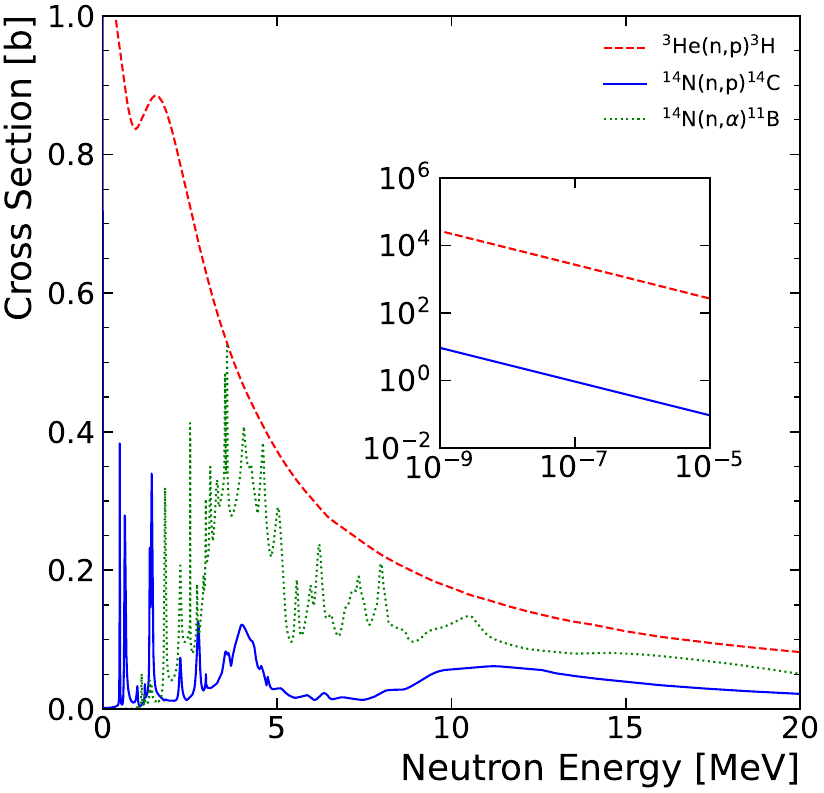}
  \vspace{-0.3cm}
\caption{Neutron capture cross-section  for different elements.\label{fig:neutronCrossSection}}
\vspace{-0.45cm}
\end{figure}
Together the two reactions have cross-sections comparable to the \ce{^3He(n,p)^3H} for fast neutrons of energies above \SI{2}{\mega\electronvolt}, as shown in Fig.~\ref{fig:neutronCrossSection}. The \ce{(n,p)} reaction also has a significant cross-section at thermal neutron energies, although it is approximately three orders of magnitude less than that of the \ce{^3He(n,p)^3H}. 

The nitrogen-filled spherical proportional counter offers several advantages, including low $\gamma$-ray sensitivity, use of an inert and inexpensive gas, and the possibility to reach large active masses through high-pressure operation, thereby suppressing the wall effect. Nitrogen, however, is a challenging operating gas for proportional counters: its first Townsend coefficient is comparatively low, requiring a high ratio of electric field strength to gas pressure ($E/P$) to achieve adequate gas gain~\cite{Knights:2025ogz}.
 
The first experimental demonstration of the method~\cite{Bougamont:2015jzx} employed a single-anode sensor and was limited to operating pressures below \SI{0.5}{bar}, with an anode voltage of approximately \SI{6.2}{\kilo\volt} required to reach sufficient gas gain. 

The advent of the \achinos\ multi-anode sensor~\cite{Giganon:2017isb,Giomataris:2020rna,Katsioulas:2022cqe} has overcome this limitation. 
The \achinos\ sensor decouples the drift and amplification electric fields by arranging several anodes at a fixed distance from the detector centre, forming a virtual sphere around a central (typically grounded) resistive electrode. As a result, the drift field in the detector volume is determined by the collective field of all the anodes, while the amplification field is set by the radius and voltage of each individual anode. This has enabled operation of a \SI{30}{\centi\meter} diameter spherical proportional counter at pressures up to \SI{1.8}{bar} of nitrogen with anode voltages below \SI{6}{\kilo\volt}, and the measurement of  thermal and fast neutrons from an \ce{^{241}Am}-\ce{^9Be} source~\cite{Giomataris:2022bvz}. Subsequent measurements at the University of Birmingham MC40 cyclotron extended the range to fast neutrons with energies up to \SI{8}{\mega\electronvolt}, produced through the \ce{^9Be(d,n)^{10}B} reaction~\cite{Giomataris:2022kxw}. 

A further development has been the implementation of individual anode read-out~\cite{Herd:2023hmu}. This enabled anode-by-anode gain calibration, resulting in improved energy resolution 
and opening the way to event localisation and track reconstruction.

\section{Characterisation of the \achinos\ multi-anode sensor}
\label{sec:achinos}
 
The sensor used in the present measurements is an 11-anode \achinos, with stainless steel, \SI{1}{\milli\meter} in diameter, spherical anodes arranged on the vertices of a truncated  icosahedron. The twelfth vertex is occupied by the grounded support rod. The anodes are at 
$r_A = \SI{1.4}{\centi\meter}$ 
from the detector centre. The central structure is additively manufactured, and then coated with a layer of Diamond-Like Carbon (DLC) 
using magnetron sputtering~\cite{Giomataris:2020rna}. 

DLC is a type of amorphous carbon containing both the diamond and the graphite crystalline phase. Thanks to its resistivity, in addition to structural, chemical and thermal stability, DLC coating 
offers a novel method for producing high quality resistive materials for gaseous
detectors, providing stable operation against discharges. 
The measured resistance between antidiametric points on the surface is \SI{10}{\giga\ohm}. The \achinos\ structure was mounted on a copper rod
and electrically connected to it using a conductive Araldite-copper (70\%:30\%) mixture. 

The anode radius and their distance from the central structure are crucial for achieving a homogeneous response, as discussed in Ref.~\cite{Herd:2023hmu}. To mitigate variations in these parameters, high-quality anode spheres (grade 10) were employed, featuring a sphericity, diameter tolerance, and lot variation of \SI{0.25}{\micro\meter}.
Furthermore, the distance between the central structure and the anodes was adjusted using a digital microscope, enabling a positioning precision of \SI{10}{\micro\meter}. 
Silver epoxy was used to attach the anode spheres to the wires, providing a well-defined resistivity of \SI{1.8e-3}{\ohm\centi\meter}, which is an improvement over the approximate \SI{e-2}{\ohm\centi\meter} resistivity of the Araldite/copper powder (70\%:30\%) mixture previously employed.  

\begin{figure}[htbp]
\centering
\includegraphics[width=0.49\linewidth]{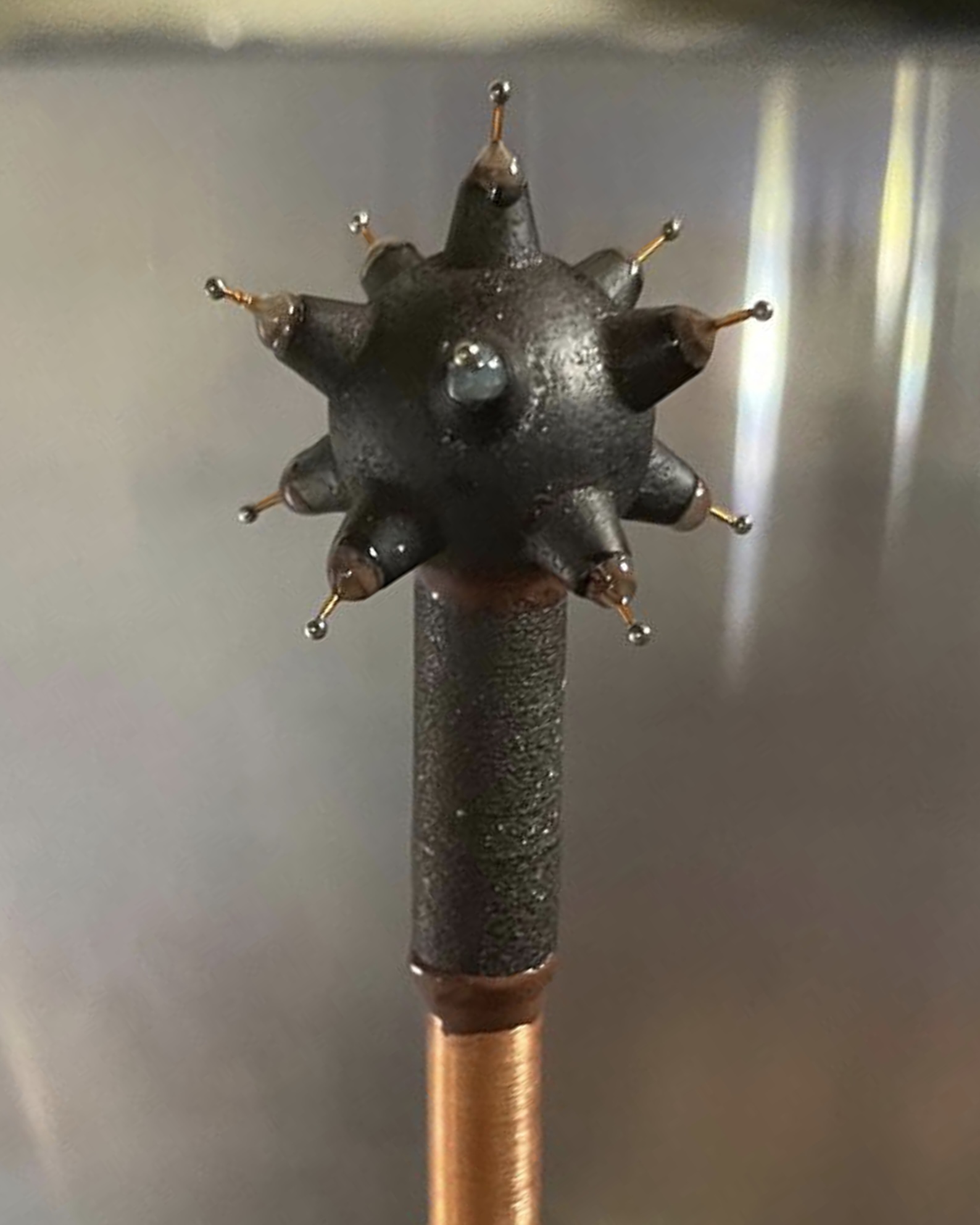}
\caption{The 11-anode \achinos\ sensor, with \SI{1}{\milli\meter} in diameter anodes placed at \SI{1.4}{\centi\meter} from the detector centre.\label{fig:achinos}}
\end{figure}

The prepared sensor is shown in Fig.~\ref{fig:achinos}. The anodes are classified as far~(F) or near~(N) based on their distance from the grounded support rod. Each anode is assigned a label used to identify it throughout the readout chain. The five anodes closest to the rod are labelled N1 to N5, the anode located diametrically opposite the rod is labelled F0, and the remaining anodes are labelled F1 to F5.

The \achinos\ employed in this work features individual read-out of each anode. 
Each of the eleven anodes is connected through its own wire to a dedicated read-out channel based on a CREMAT CR-110-R2.2 charge-sensitive preamplifier. 
The high voltage is supplied to each anode individually through the same read-out board, allowing per-anode voltage tuning. 
The V1725S 14-bit  analogue-to-digital converter (ADC) constructed by CAEN  is used for signal digitisation, providing 16 channels with a dynamic range of \SI{2}{\volt}. 
The digitiser allows for a maximum sampling frequency of \SI{250}{\mega\sample\per\second}, but is operated at \SI{125}{\mega\sample\per\second}.
The digitised signal is sent via the CAEN A4818 CONET2 Adapter 
to the data acquisition PC operating the CAEN WaveDump2  software.

The sensor was characterised in February 2026 using N$_2$ and Ar/CH$_4$ (98\%:2\%) at \SI{1}{bar}, for direct comparison with previous measurements~\cite{Giomataris:2020rna}. The characterisation was performed using a \ce{^{55}Fe} source, emitting X-rays of \SI{5.9}{\kilo\electronvolt}, and a \ce{^{241}Am} source, emitting $\alpha$-particles with an energy of \SI{5.49}{\mega\electronvolt}. The source was placed on the inner surface of the cathode and could be repositioned without opening the detector vessel, enabling the individual calibration of each anode~\cite{Herd:2023hmu}.

\begin{figure}[htb]
\centering
\vspace{-0.3cm}
\subfigure[\label{fig:ACHINOS_Fe}]{\includegraphics[width=0.49\linewidth]{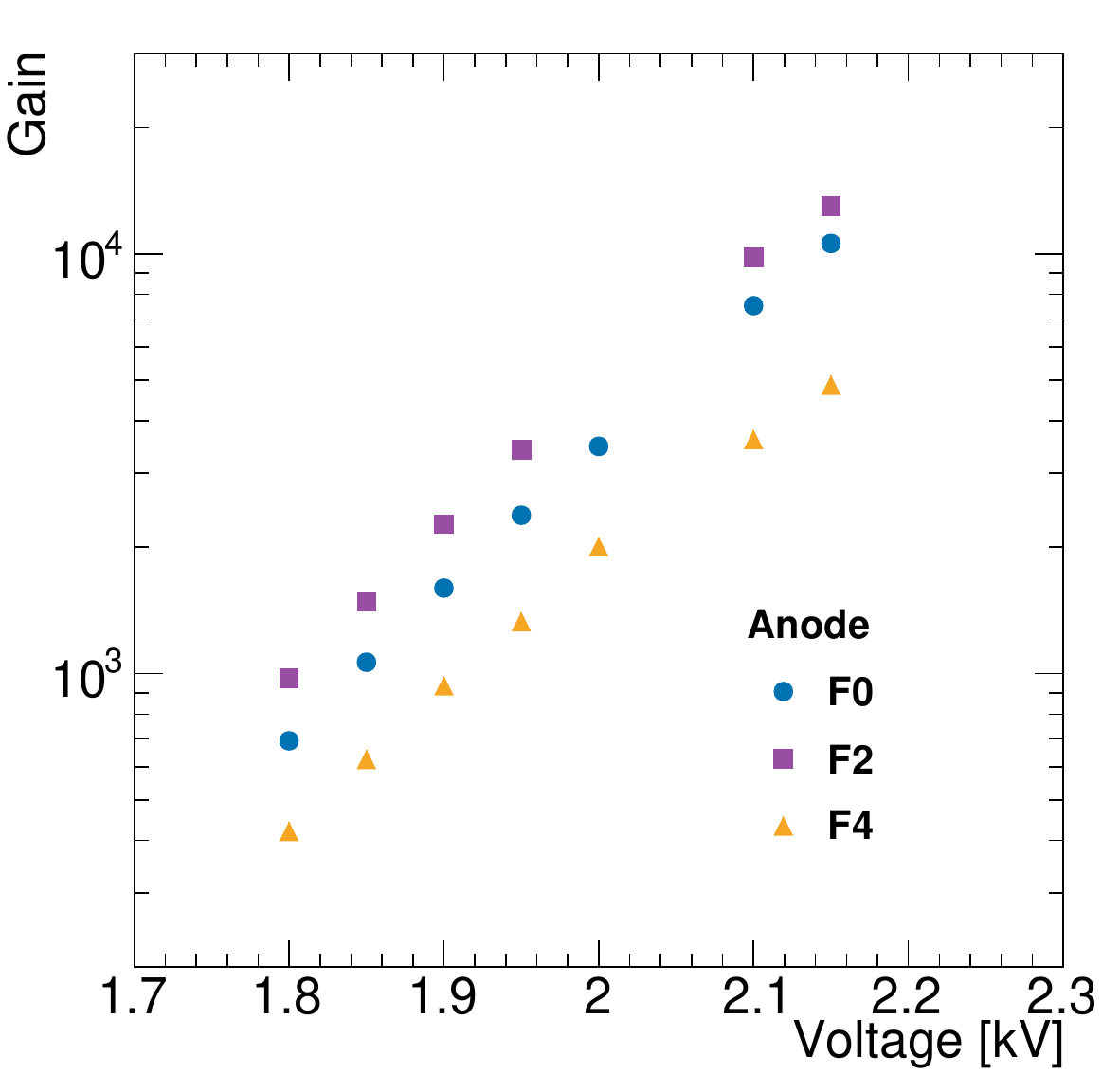}}
\subfigure[\label{fig:ACHINOS_Am}]{\includegraphics[width=0.49\linewidth]{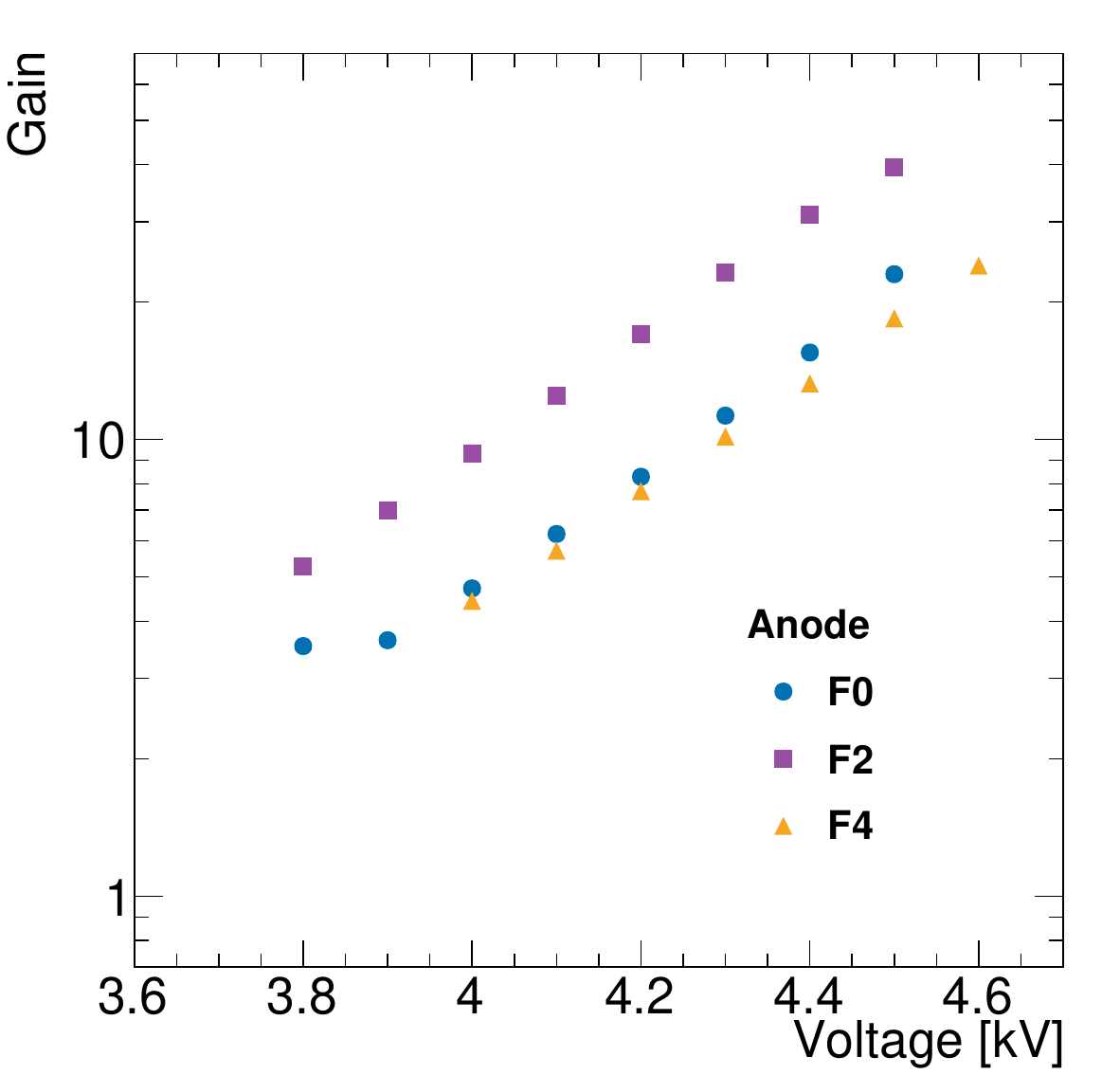}}
\caption{Gain versus voltage curve for three anodes: \subref{fig:ACHINOS_Fe} \SI{1}{bar} Ar/CH$_4$ (98\%:2\%) using \ce{^{55}Fe}; and \subref{fig:ACHINOS_Am} \SI{1}{bar} nitrogen using \ce{^{241}Am}. 
\label{fig:ACHINOS_gain}}
\end{figure}

For each anode, the pulse amplitude distribution was recorded and fitted with a normal distribution, yielding the mean response and local energy resolution. As an example, the gain as a function of bias voltage for three anodes is presented in Fig.~\ref{fig:ACHINOS_Fe}. The gains of all other anodes were found to lie between the curves shown. The observed trend confirms the expected exponential dependence, in agreement with previous studies~\cite{Giomataris:2020rna}. 
The corresponding gain curve for the \achinos\ sensor operated in N$_2$ is shown in Fig.~\ref{fig:ACHINOS_Am}. 
The relative differences of the anode mean response was subsequently used to equalise the gain, either by applying per-anode  order \SI{1}{\percent} voltage adjustments~\cite{Herd:2023hmu}, or by appropriately scaling the recorded signals during offline analysis. 

\begin{figure}[htb]
\centering
\vspace{-0.6cm}
\subfigure[\label{fig:ACHINOS_Fe2_a}]{\includegraphics[width=0.49\linewidth]{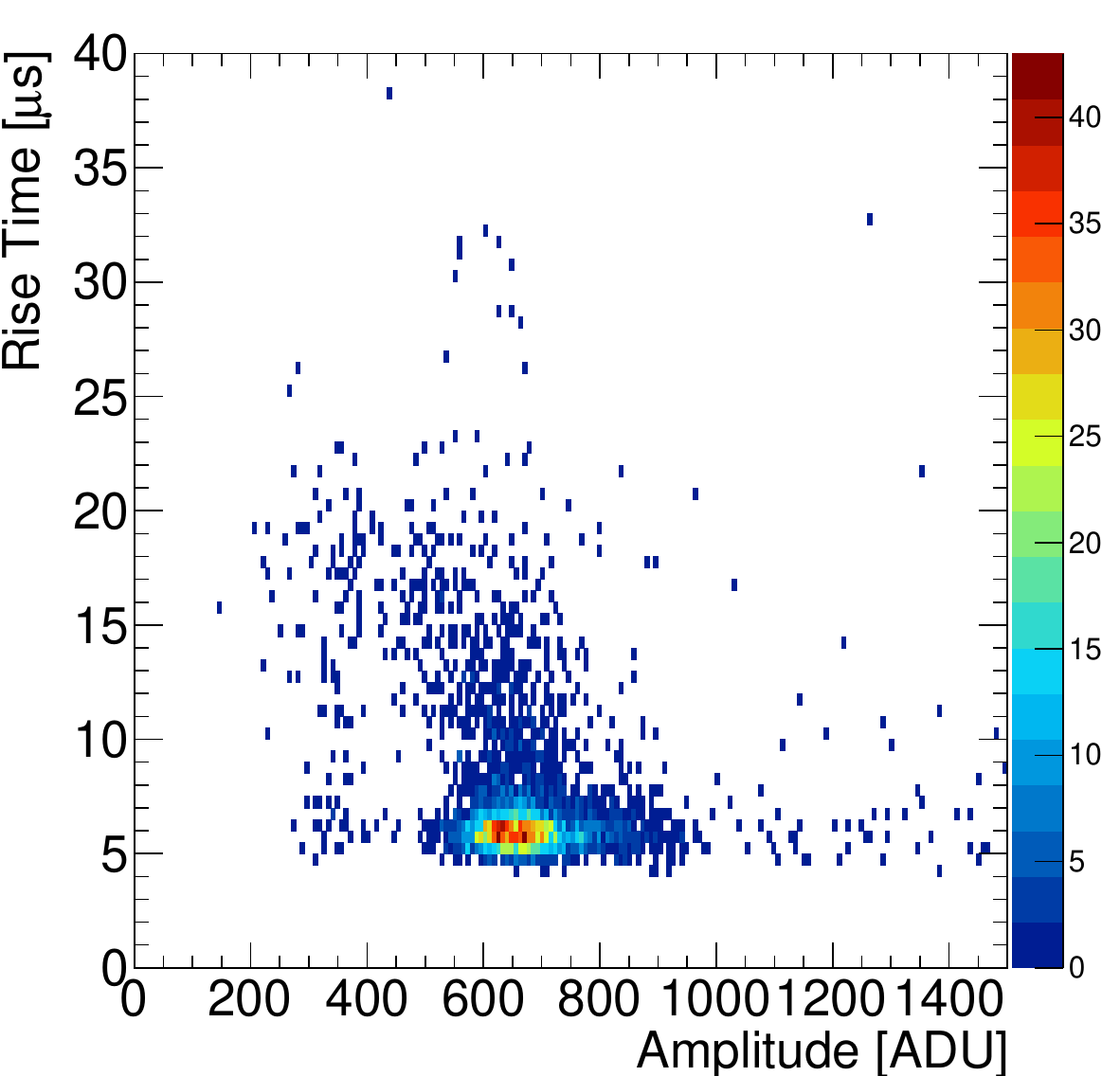}}
\subfigure[\label{fig:ACHINOS_Fe2_b}]{\includegraphics[width=0.49\linewidth]{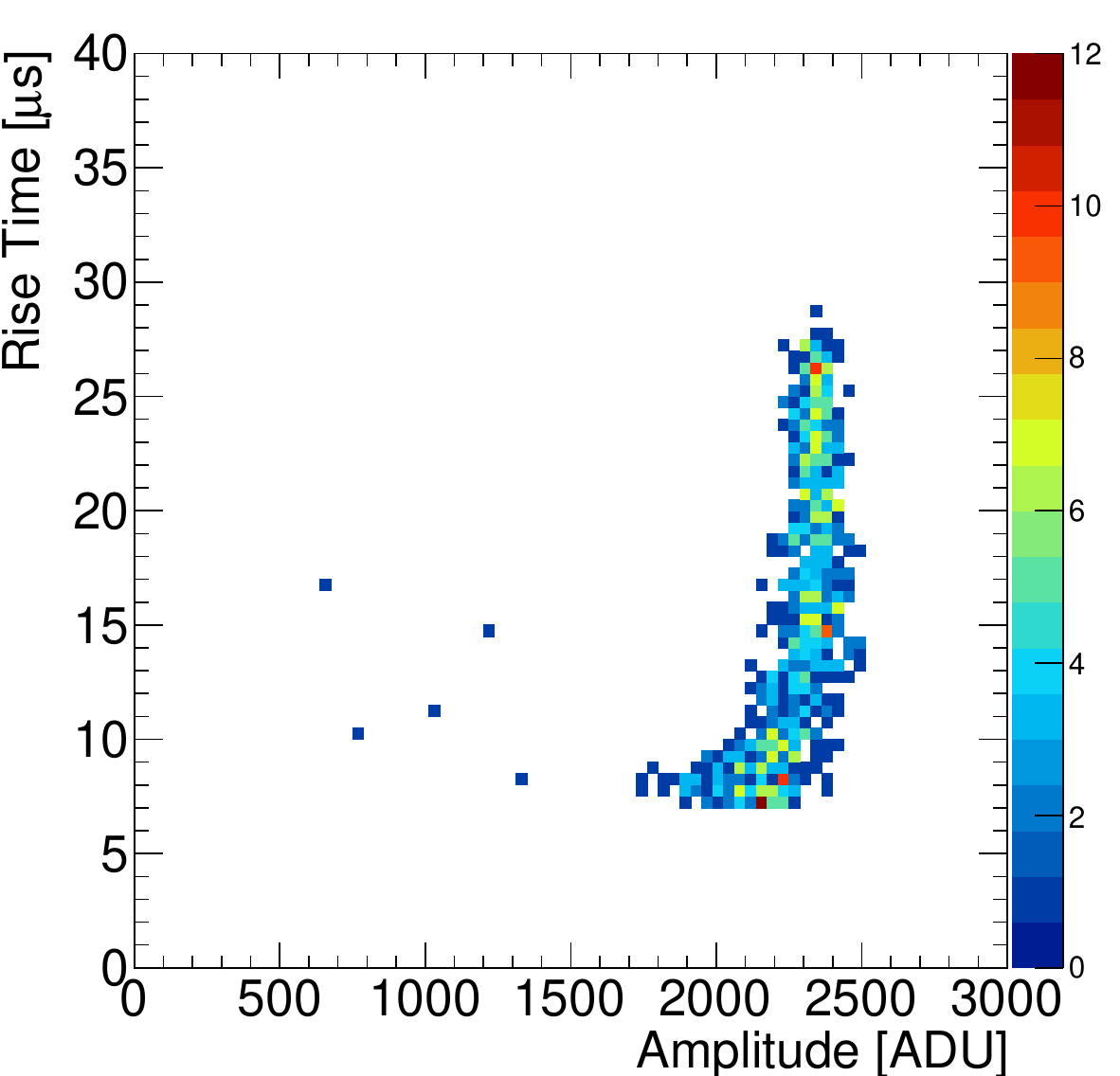}}
\caption{F0 distribution for rise time versus amplitude with \subref{fig:ACHINOS_Fe2_a} $^{55}$Fe in Ar/CH$_4$ at \SI{1.9}{\kilo\volt}  and \subref{fig:ACHINOS_Fe2_b} Am in N$_2$ at \SI{4.2}{\kilo\volt}.\label{fig:ACHINOS_Fe2}}
\end{figure}

An example of the pulse peak amplitude versus pulse rise time, defined as the time required for the signal to rise from \SI{10}{\percent} to \SI{90}{\percent} of its amplitude, is shown for anode F0 in Fig.~\ref{fig:ACHINOS_Fe2_a}  \SI{5.9}{\kilo\electronvolt} X-rays in Ar/CH$_4$  and in Fig.~\ref{fig:ACHINOS_Fe2_b} for \SI{5.49}{\mega\electronvolt} $\alpha$-particles in N$_2$. 
The two source populations exhibit distinct distributions in the pulse-amplitude--rise time plane, providing additional means of event discrimination and background rejection.


The stability of the gain was investigated by monitoring the detector response to $\alpha$-particles over time. Following the ramping-up of the high voltage, the gain was observed to increase, reaching a plateau after approximately \SI{18}{\hour} of continuous operation. The origin of this behaviour is under investigation. Although this effect introduces a delay before the detector reaches optimal operating conditions, the gain remains stable once equilibrium is achieved. Moreover, subsequent small variations in bias voltage do not seem to affect the gain stability.

\section{Neutron beam and experimental setup}
\label{sec:setup}

The Tandem Accelerator Laboratory is a research infrastructure of the
Institute of Nuclear and Particle Physics of NCSR ``Demokritos''~\cite{Harissopulos:2021lgh}. 
It features a \SI{5.5}{\mega\volt} T11/25 Van de Graaff Tandem accelerator 
manufactured by the High Voltage Engineering Corporation (HVEC), shipped to ``Demokritos'' in 1971
and delivering its first beam in 1973. Through a series of
upgrades~\cite{Lagoyannis:2023mnv}, the accelerator has remained a cornerstone of
nuclear research at NCSR ``Demokritos'', supporting a broad programme that spans
nuclear structure and nuclear astrophysics as well as ion-beam analysis and
X-ray fluorescence spectrometry (XRF) for the study of problems of societal impact.

\begin{figure*}[h!]
\centering
\includegraphics[width=0.395\linewidth]{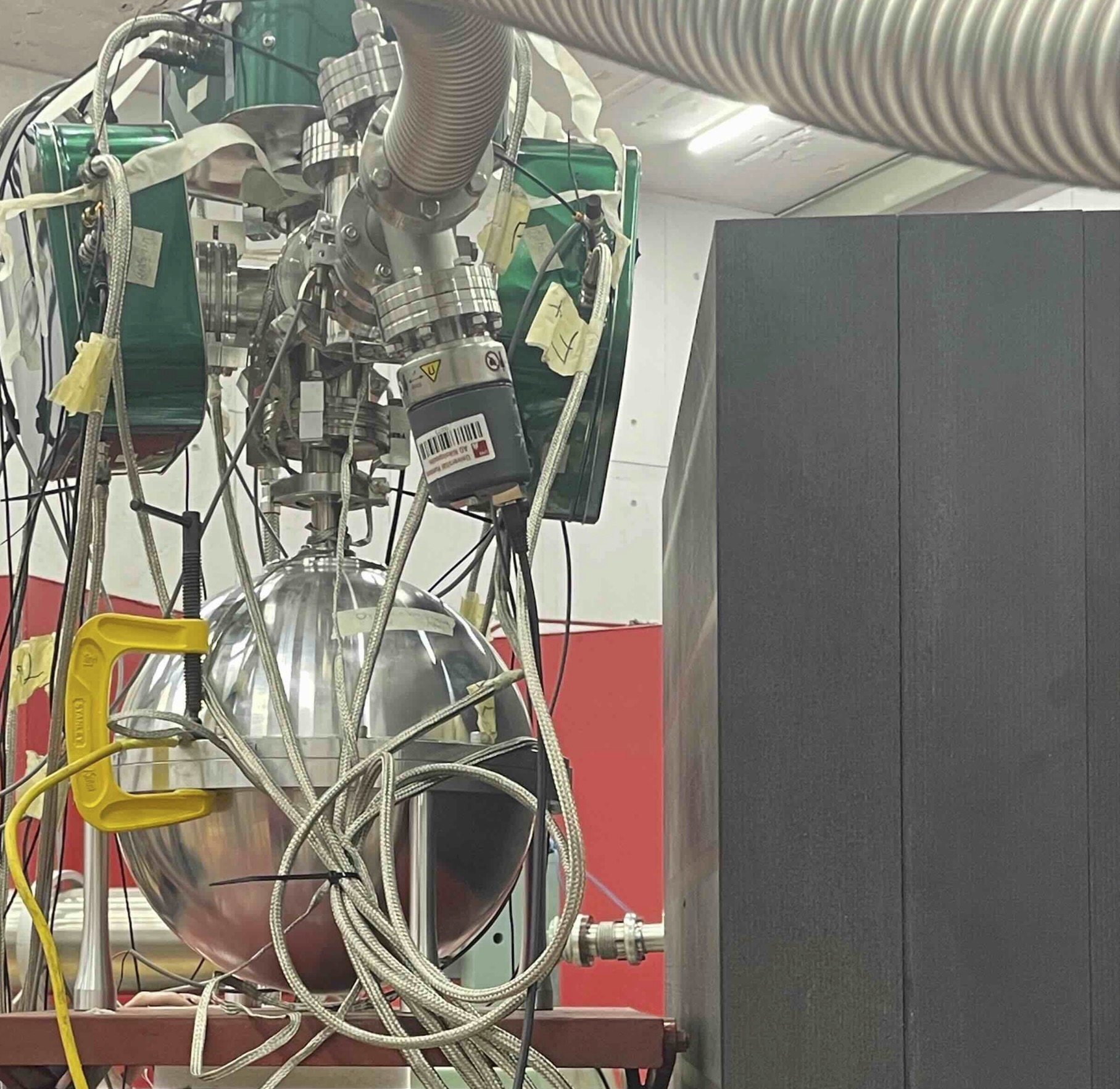}
\includegraphics[width=0.35\linewidth]{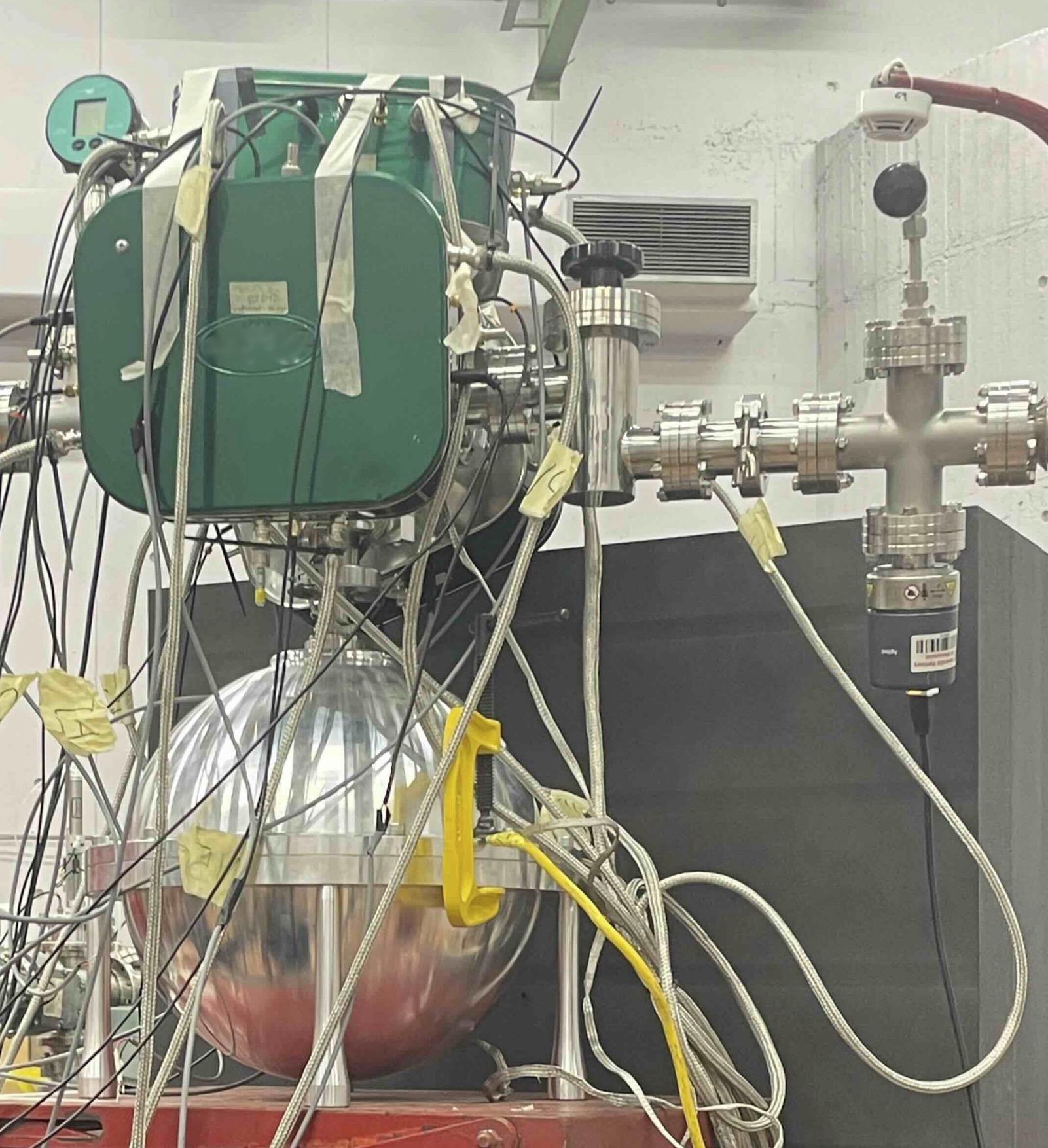}

\caption{The nitrogen-filled spherical proportional counter installed for data-taking, shown from two viewpoints. The
detector and its gas-system and read-out hardware (mounted above) sit
immediately downstream of the borated-HDPE shielding castle (dark block) enclosing
the \ce{LiF} target. The proton beam reaches the target through the beamline
entering from the right on the left-hand view.
\label{fig:setup}}
\end{figure*}

The accelerator is housed on the ground floor of the building, confined by
borated concrete walls of up to \SI{1.2}{\meter} thickness. The beam is
guided into two target rooms, referred to as ``Green'' and ``Red'', where the
experiments are installed. The facility
provides mono-energetic neutron beams through several reactions, with fluxes of
order \SIrange{e5}{e6}{\per\square\centi\meter\per\second}~\cite{Harissopulos:2021lgh}.
Neutrons in the energy ranges tens of \si{\kilo\electronvolt} to \qty{4}{\mega\electronvolt}, \SIrange{4}{11.5}{\mega\electronvolt}
and \SIrange{16}{20.5}{\mega\electronvolt} are delivered via
the \ce{^7Li(p,n)^7Be}, \ce{^2H(d,n)^3He} and
\ce{^3H(d,n)^4He} reactions, respectively~\cite{Harissopulos:2021lgh}.

In the measurements reported here, the interest is in the lower part of the offered energy range,
and for this reason the \ce{^7Li(p,n)^7Be} reaction is employed on a lithium fluoride (LiF) target. Above
its threshold of $E_p = \SI{1.881}{\mega\electronvolt}$ this reaction produces a mono-energetic
neutron group, with the neutron energy set by the proton energy and the
emission angle. For proton energies above approximately $\SI{2.37}{\mega\electronvolt}$
a second, lower-energy neutron group appears, corresponding to the
\SI{429}{\kilo\electronvolt} first excited state of \ce{^7Be}. The present study
covers neutron energies of \SIrange{0.75}{2.75}{\mega\electronvolt}.
The fraction of the second neutron group in the overall population remains below  \qty{20}{\percent}, and is maximum at approximately $E_p=\SI{3}{\mega\electronvolt}$.

The experimental setup was installed at the
``Red'' experimental area in the \ang{45} beamline and shown in
Fig.~\ref{fig:setup}, with the detector configuration and data acquisition chain described in Section~\ref{sec:achinos}.
The detector was operated at 
\SI{1}{bar} gas pressure and a voltage of \SI{4.2}{\kilo\volt}. This provided a
gas gain adequate to accommodate the full range of neutron-induced energy
depositions within the dynamic range of the electronics chain. Continuous in-situ calibration was provided by an \ce{^{241}Am} $\alpha$-particle source placed on the inner surface of the cathode. 

To suppress backgrounds from neutron scattering in the experimental hall
and to define the beam incident on the detector, the target region was enclosed in
a castle of borated high-density polyethylene (HDPE) with a boron content of
\SI{10}{\percent} by weight, in which the hydrogen moderates fast neutrons and the
\ce{^{10}B} subsequently captures the thermalised neutrons.
The castle comprised ten slabs, each of
dimensions
$\SI{75}{\centi\meter}\times\SI{75}{\centi\meter}\times\SI{10}{\centi\meter}$.
On the downstream side, four slabs, each with a small approximately central
circular aperture of  \SI{8}{\milli\meter} diameter, were stacked with the face
closest to the target positioned approximately \SI{10}{\centi\meter} from the
\ce{LiF} target; their aligned apertures formed a channel through which a
collimated neutron beam reached the detector. The remaining six slabs, each with a
square central aperture of
$\SI{15}{\centi\meter}\times\SI{15}{\centi\meter}$, formed the upstream and
lateral walls, with the beam-entry opening on the upstream side closed off by a
\SI{10}{\centi\meter}-thick layer of the same material. The spherical proportional
counter was positioned immediately downstream the shielding assembly. In the right-handed coordinate system with the target at its origin,
 x-axis being the beam direction, which is practically horizontal, z-axis being vertical with upwards direction, and the y-axis selected to complete the right-handed system, the centre of the detector was at $\left(\SI{777.3}{\milli\meter}, -\SI{18.2}{\milli\meter}, \SI{76.3}{\milli\meter}\right)$. These positions are known with better than \qty{1}{\milli\metre} surveying accuracy.

\section{Data analysis and results}
\label{sec:results}
 
The data collection took place at Athens in March 2026. The following seven neutron energies were used: 
$E_n =$ \SIlist[list-units=single]{0.750;1.000;1.425;1.800;2.220;2.517;2.750}{\mega\electronvolt}. These energies were selected to span the energy range over which the \ce{^{14}N(n,p)^{14}C} and \ce{^{14}N(n,\alpha)^{11}B} cross-sections evolve and the relative contribution of the two reactions changes. The corresponding proton beam energies were estimated from the reaction kinematics and target thickness and were confirmed by dedicated simulations, with an associated uncertainty on the neutron energy of the order of \qty{10}{\kilo\electronvolt}. 

\subsection{Event selection and pulse-shape response}
\begin{figure}[htbp]
    \centering
    \includegraphics[width=0.90\linewidth]{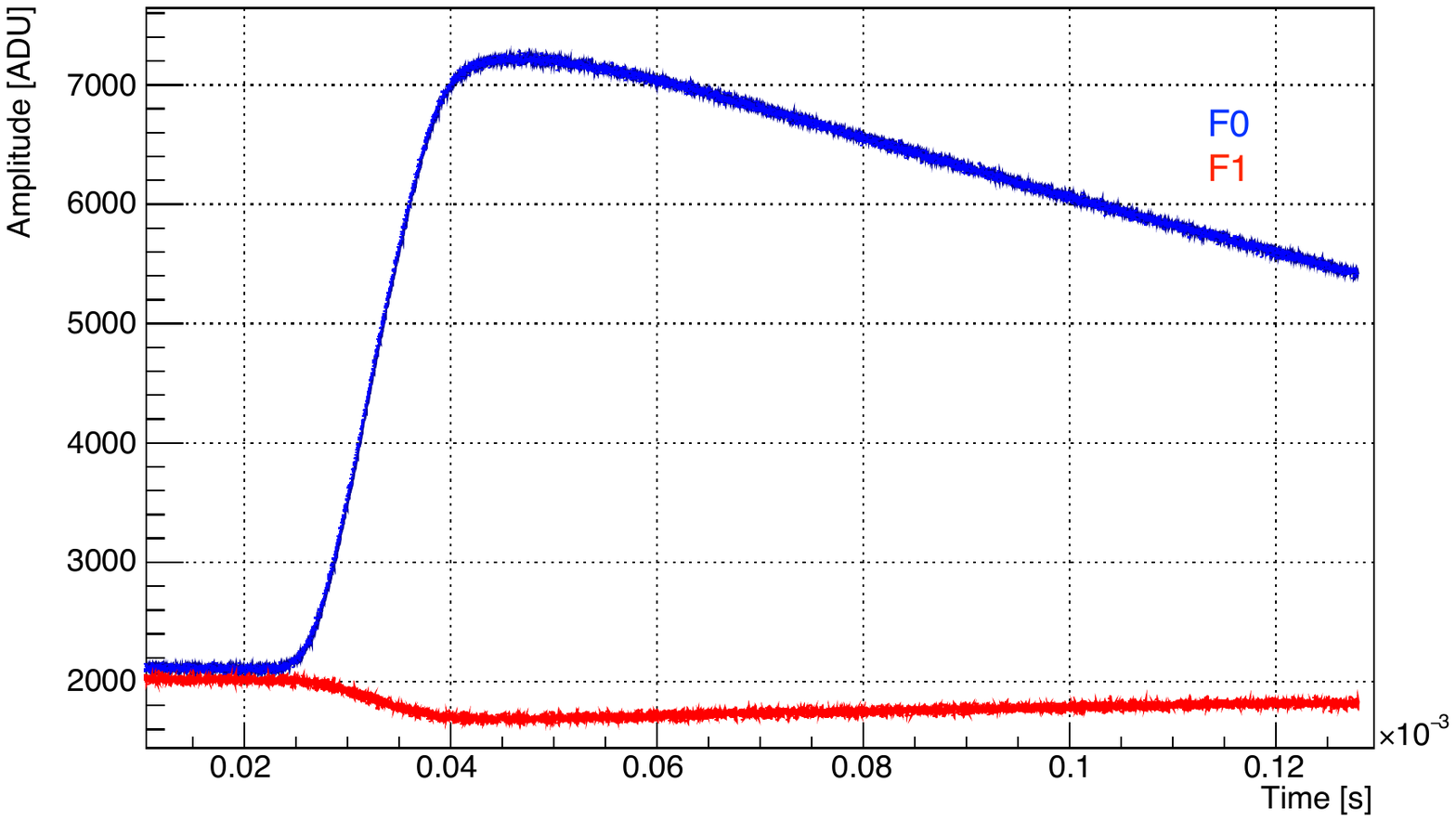}
    \caption{Typical waveforms for an event where  the avalanche develops on F0: in blue, a positive signal induced on F0, and, in red, the simultaneous negative pulse recorded on the neighbouring anode F1.
    \label{fig:Wave}}
\end{figure}

\begin{figure*}[htbp]
\centering
\subfigure[\label{fig:rva_a}]{\includegraphics[width=0.33\linewidth]{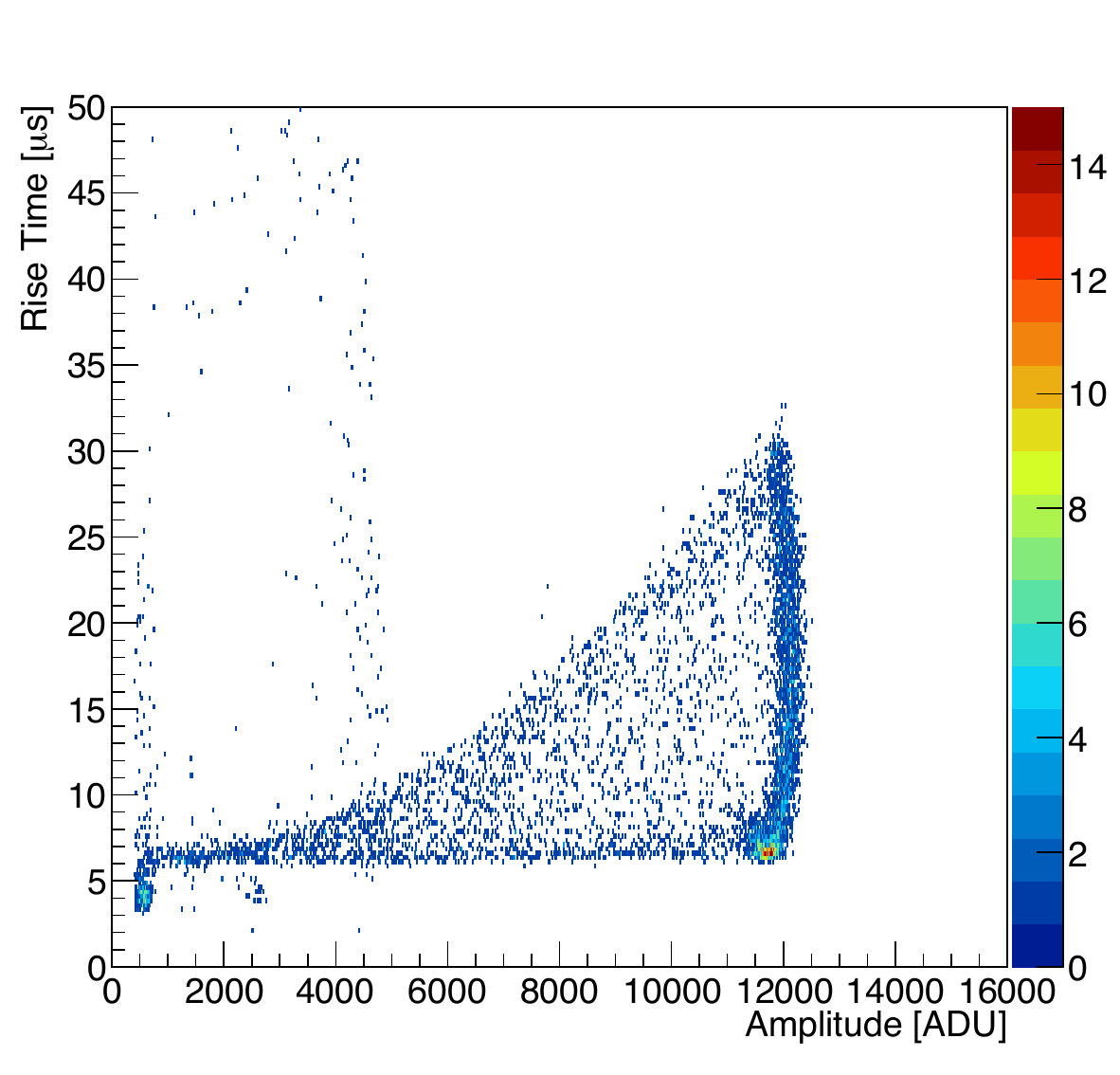}}
\subfigure[\label{fig:rva_b}]{\includegraphics[width=0.33\linewidth]{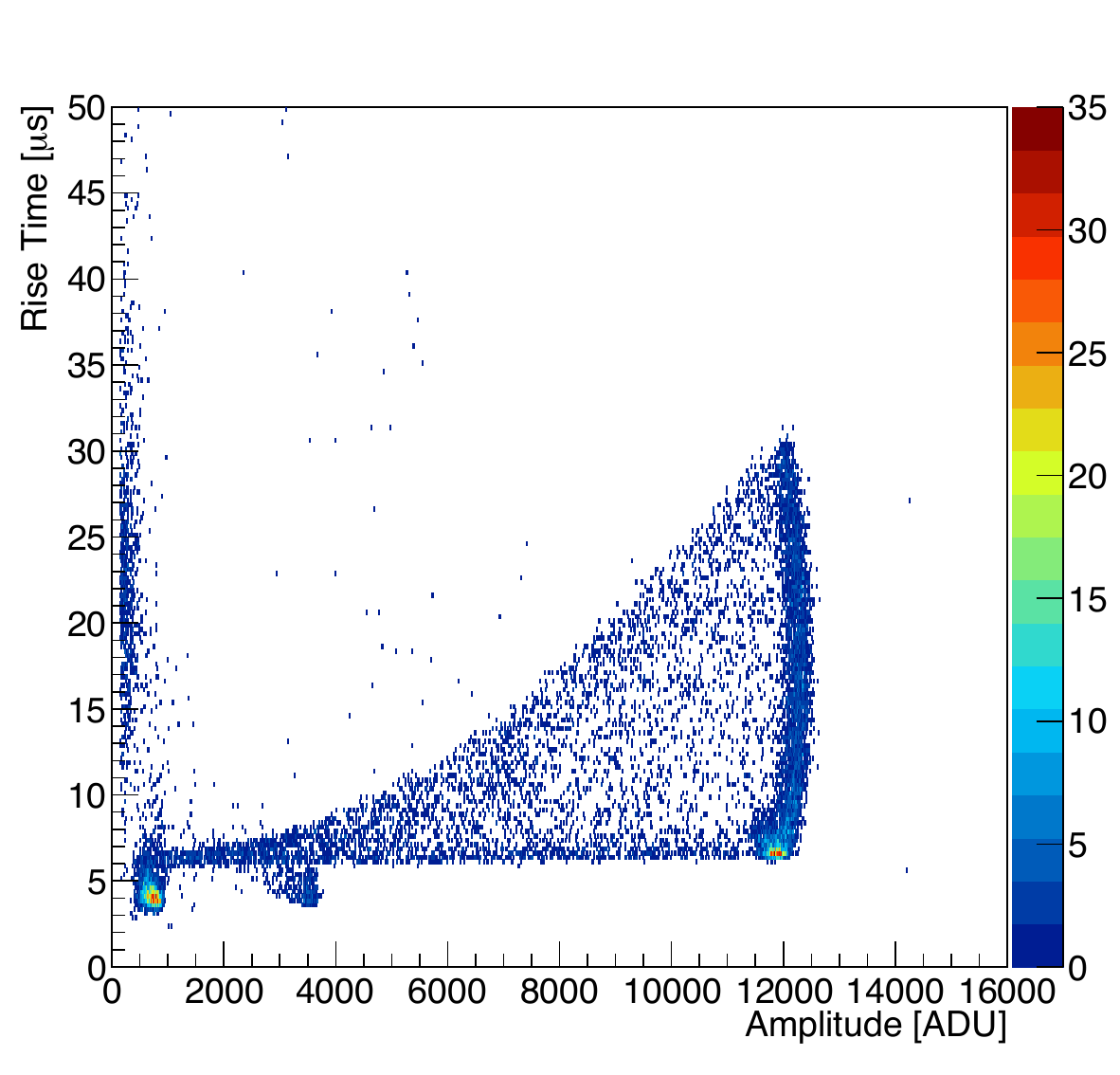}}
\subfigure[\label{fig:rva_c}]{\includegraphics[width=0.33\linewidth]{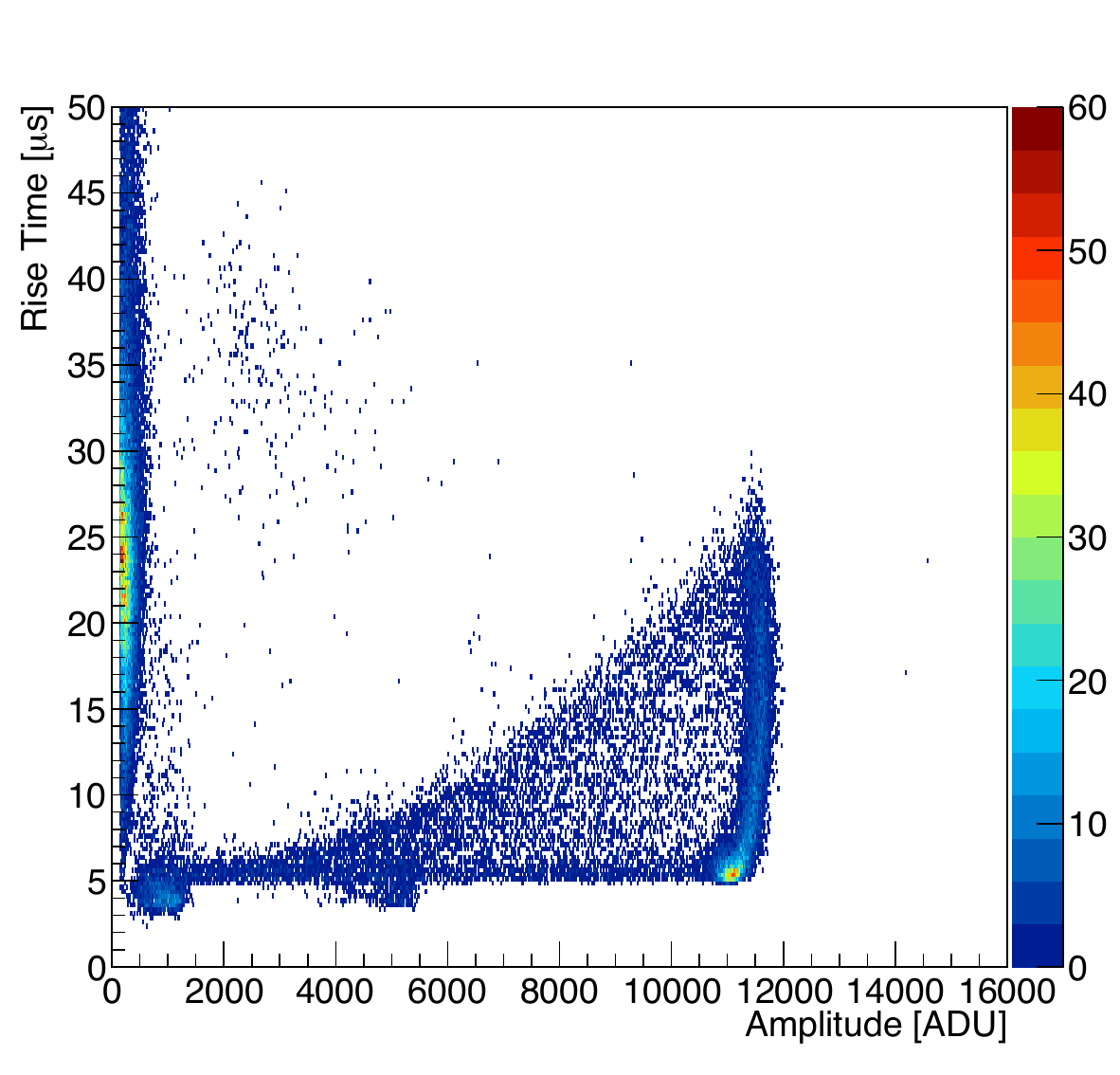}}
\caption{Pulse rise time versus pulse amplitude for the measurement at $E_n = \SI{1.425}{\mega\electronvolt}$ \subref{fig:rva_a}. The population at rise times from \SIrange{13}{50}{\micro\second} and amplitude at approximately \SI{4500}{\ADU} correspond to the \ce{^{14}N(n,p)^{14}C} reaction, while the population at low rise time and amplitude at \SI{2500}{\ADU} corresponds to the \ce{^{14}N(n,\alpha)^{11}B} reaction. Also shown are the $\SI{1.800}{\mega\electronvolt}$ distribution \subref{fig:rva_b}, with corresponding populations at around \SI{3500}{\ADU} and \SI{4500}{\ADU} and $\SI{2.750}{\mega\electronvolt}$ distributions \subref{fig:rva_c} with corresponding populations at \SI{5000}{\ADU} and \SI{6500}{\ADU} \label{fig:risetime_vs_amp}}
\end{figure*}

Figure \ref{fig:Wave} shows a typical pulse recorded when the avalanche develops close to anode F0. The signal induced on F0 is positive, while the neighbouring anode F1 simultaneously records a negative pulse~\cite{Katsioulas:2022cqe}.
The pulse rise time and pulse amplitude are used as the primary pulse-shape parameters for event characterisation. 
As the beam was directed in the vicinity of F0, the trigger was based exclusively on that anode. 
Accordingly, the analysis is restricted to events in which all ionisation electrons are collected on F0, as indicated by a positive signal observed exclusively on that anode. Thanks to the independent read-out of the individual anodes, this can be achieved by requiring that all other anodes do not record a significant signal. 
An additional lower threshold of \SI{2}{\micro\second} on the pulse rise time was applied to select physical events.
Figure~\ref{fig:risetime_vs_amp} shows the two-dimensional distribution of pulse rise time versus pulse amplitude for neutron beams of \SIlist[list-units=single]{1.425;1.800;2.750}{\mega\electronvolt}. The correlation between these two observables allows neutron-induced events to be discriminated from background and from $\alpha$-particles emitted by the calibration source. In particular, the \ce{^{14}N(n,p)^{14}C} and \ce{^{14}N(n,\alpha)^{11}B} reactions occupy distinct regions of the rise time--amplitude parameter space.
Events attributed to the \ce{^{14}N(n,\alpha)^{11}B} form a  localised distribution in both rise time and amplitude. In contrast, events attributed to the \ce{^{14}N(n,p)^{14}C} reaction populate a region with a fixed amplitude but a large spread in the rise time, as a result of their larger range. For both reactions, the corresponding pulse amplitude depends on the neutron energy. The $\alpha$-particle events from the \ce{^{241}Am} source extend over a broader amplitude range, reflecting the geometry and spatial distribution of the source and its collimator relative to the sensor.
The second, lower-energy, group of neutrons discussed in Section~\ref{sec:setup} has not been observed within the available statistics. 

\begin{figure*}[htbp]
\centering
\subfigure[\label{fig:sharedWave}]{\includegraphics[width=0.45\linewidth]{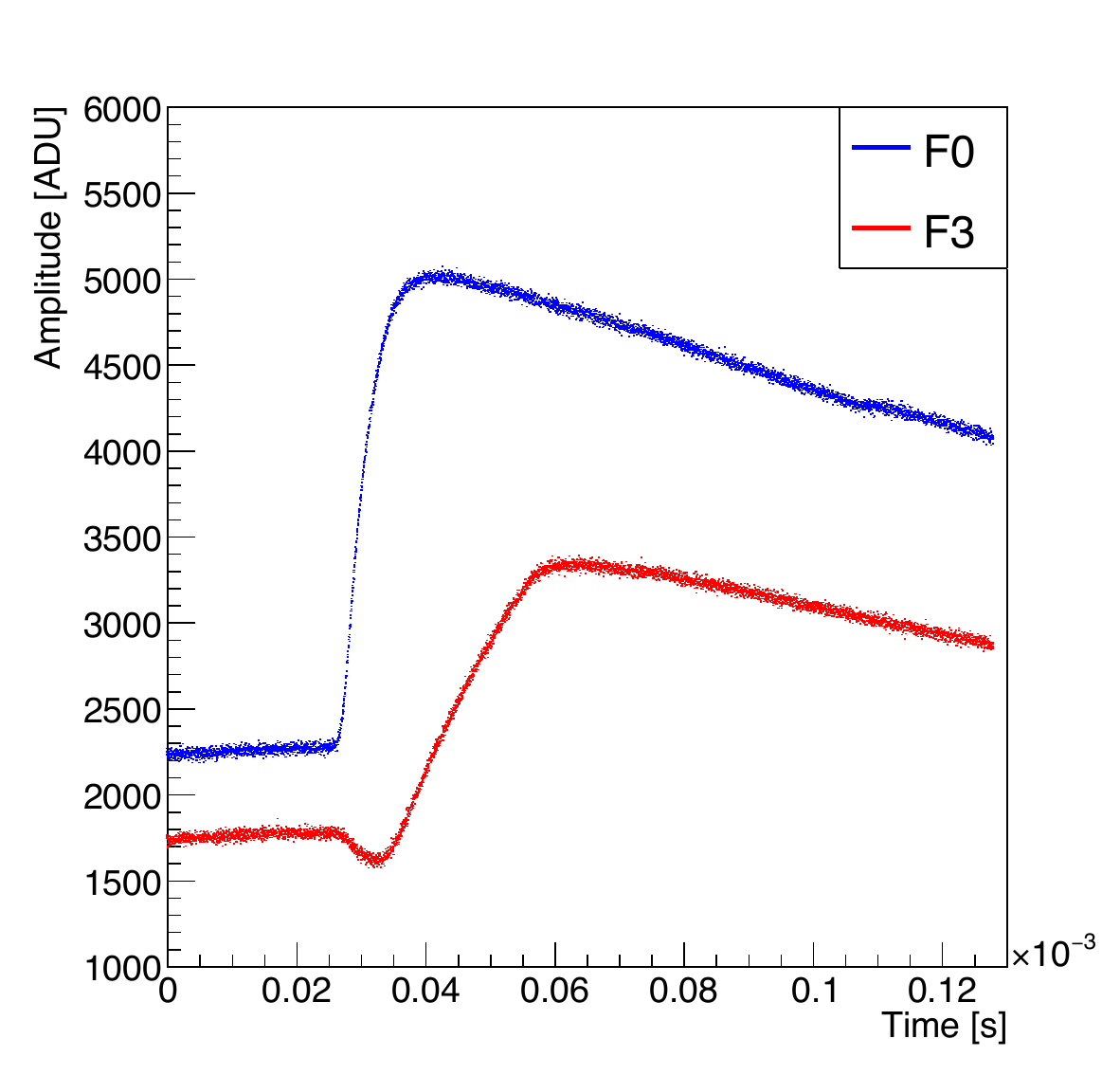}}
\subfigure[\label{fig:sharedE2}]{\includegraphics[width=0.41\linewidth]{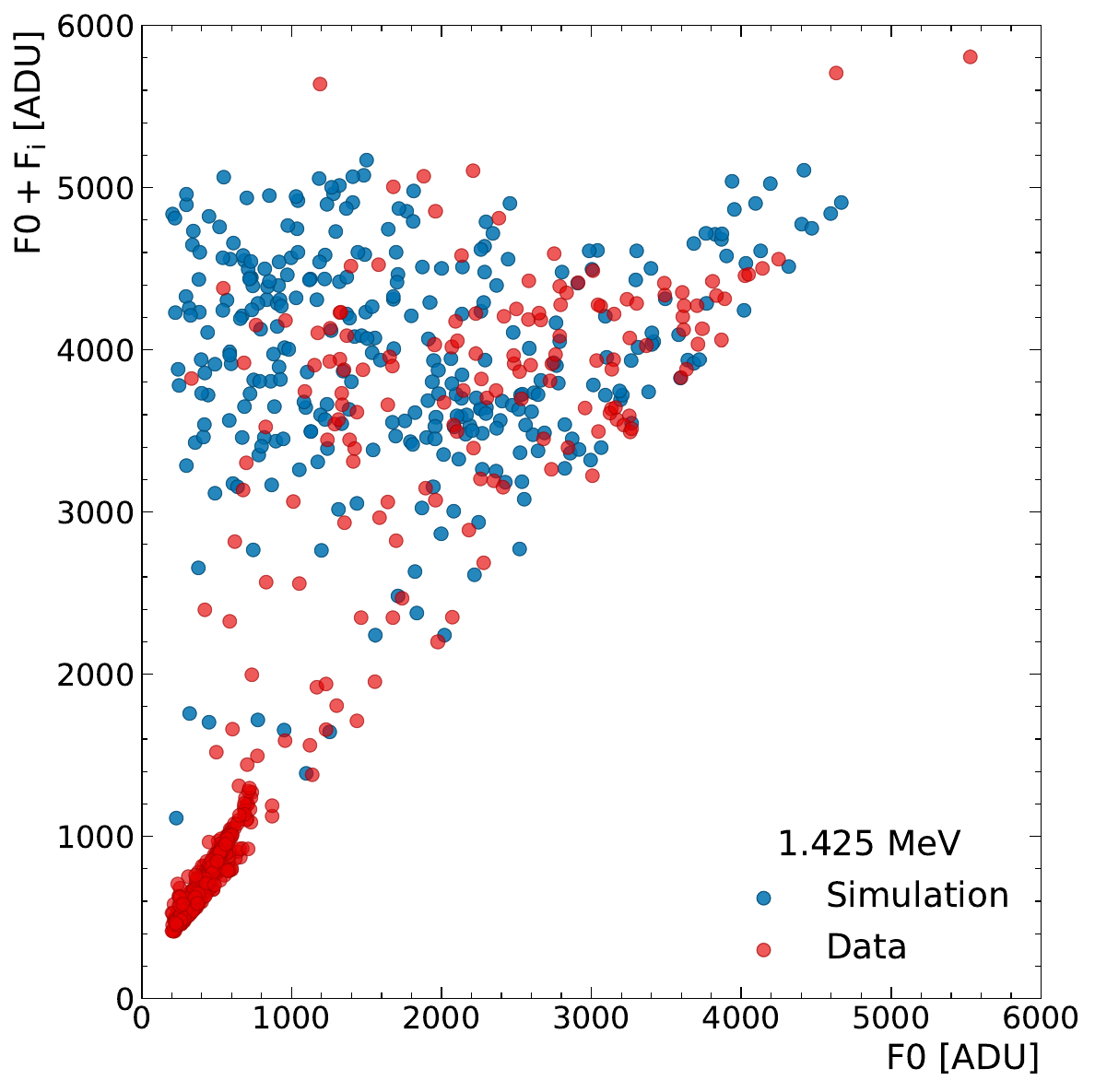}}
\caption{\subref{fig:sharedWave} Waveforms for a shared event. \subref{fig:sharedE2} Sum of amplitudes of two-anode shared event versus the amplitude of F0.\label{fig:share}}
\end{figure*}

Shared events are defined as events in which ionisation electrons arrive on two or more anodes. An example of a shared event between F0 and F3 is shown in Fig.~\ref{fig:sharedWave}. Initially, electrons arriving at F0 create an avalanche, which induces a positive signal on F0 and a corresponding negative signal on F3. Subsequently, electrons arriving at F3 create an avalanche there and induce a positive signal on that anode~\cite{Katsioulas:2022cqe}. For the  preliminary study of shared events presented, a selection is applied to retain events containing exactly two positive pulses with a time difference of less than \SI{60}{\micro\second}. In Fig.~\ref{fig:sharedE2}, the sum of the amplitudes measured on the two anodes is shown as a function of the amplitude recorded on F0. The population of events with sum of amplitudes at approximately \qty{4000}{\ADU} are identified as shared \ce{^{14}N(n,p)^{14}C} reaction events. The data observed at small amplitudes are attributed to noise. 

\begin{figure}[htbp]
\centering
\subfigure[\label{fig:sim_a}]{\includegraphics[width=0.49\linewidth]{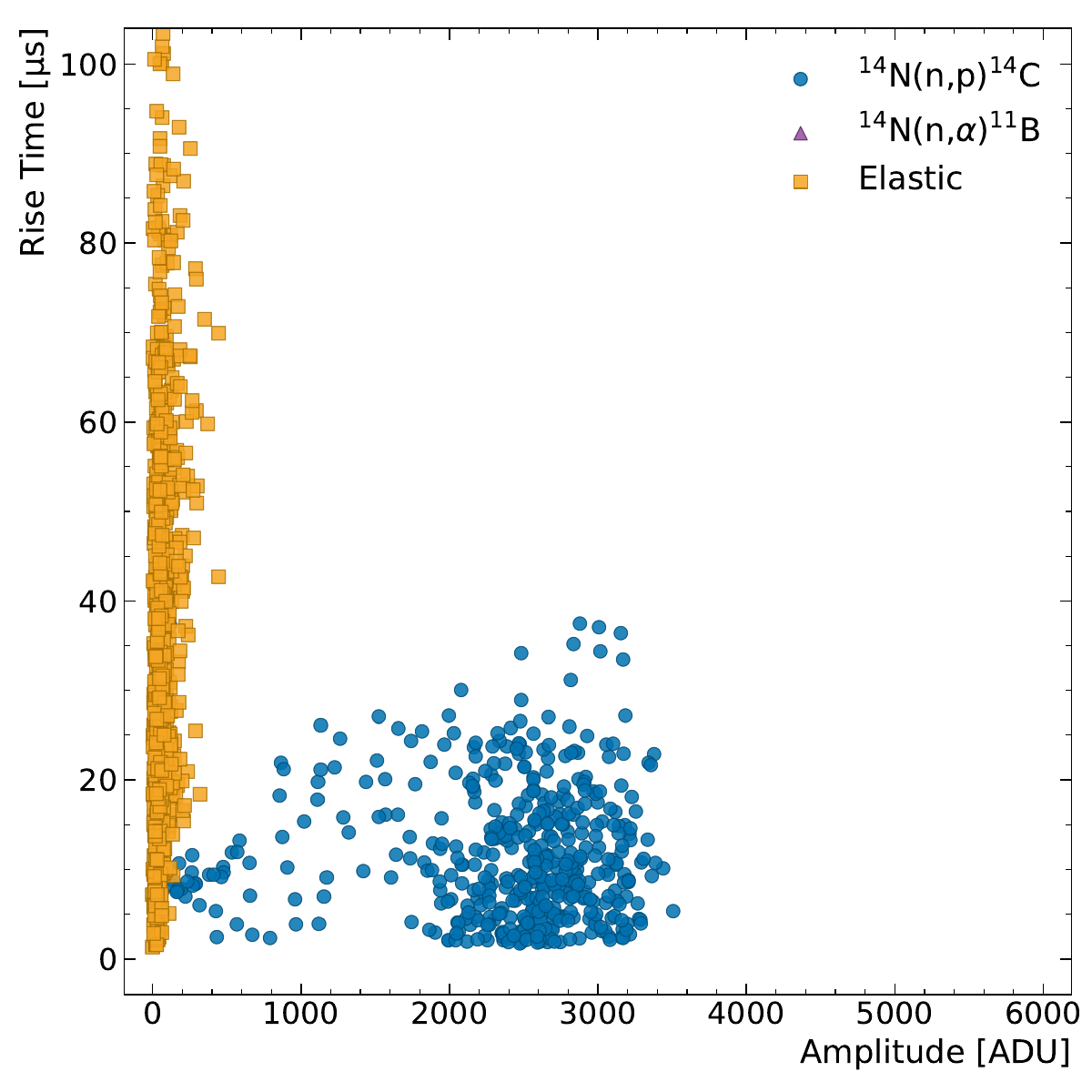}}
\subfigure[\label{fig:sim_b}]{\includegraphics[width=0.49\linewidth]{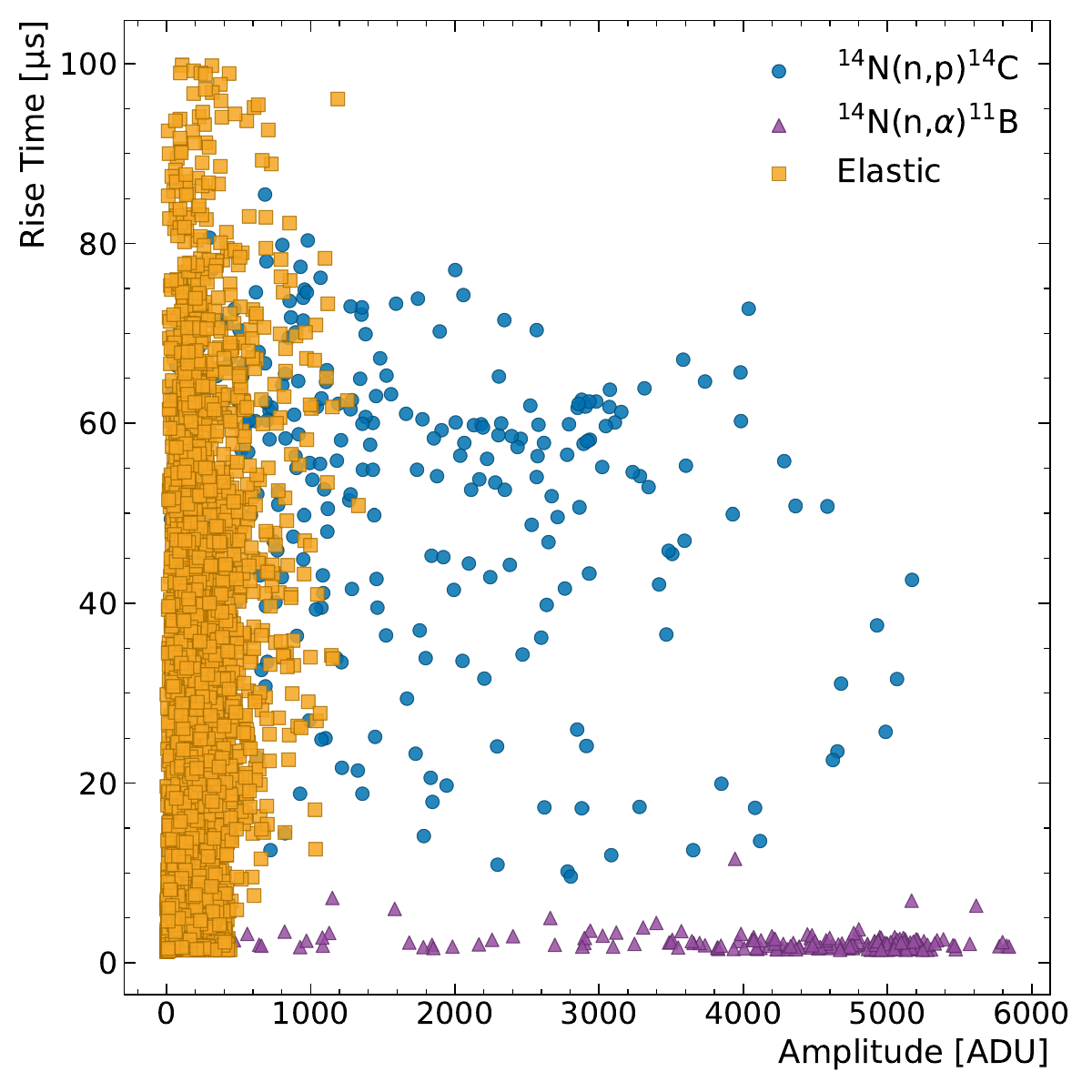}}
\caption{Simulated rise time versus amplitude distributions for neutrons of \subref{fig:sim_a} \SI{0.75}{\mega\electronvolt} and \subref{fig:sim_b} \SI{2.75}{\mega\electronvolt},  incident on the spherical proportional counter. Events corresponding to the different interactions are depicted with different markers.\label{fig:sim}}
\end{figure}

Simulations were performed for a range of incident neutron energies,  using a simulation framework~\cite{Katsioulas:2019sui} that combines Geant4~\cite{GEANT4:2002zbu} for the interactions of particles with matter, Garfield++~\cite{Veenhof:1993hz, Veenhof:1998tt} for electron transport and signal induction, and Gmsh/Elmer \cite{gmsh, elmer} for finite element calculations of the electric field. 
The amplitude and rise time were studied for the three key neutron interactions, the elastic scattering, the \ce{^{14}N(n,p)^{14}C} reaction and the \ce{^{14}N(n,\alpha)^{11}B} reaction. 
The rise time versus amplitude distributions for neutron energies of \SIlist[list-units=single]{0.75;2.75}{\mega\electronvolt} are shown in Fig.~\ref{fig:sim}, with the distinct contributions highlighted. Protons with lower amplitude than the full proton energy peak are identified as having deposited energy on the cathode, i.e. suffering wall effect.

\subsection{Energy reconstruction and uncertainties}
\begin{figure}[htbp]
    \centering
    \includegraphics[width=0.90\linewidth]{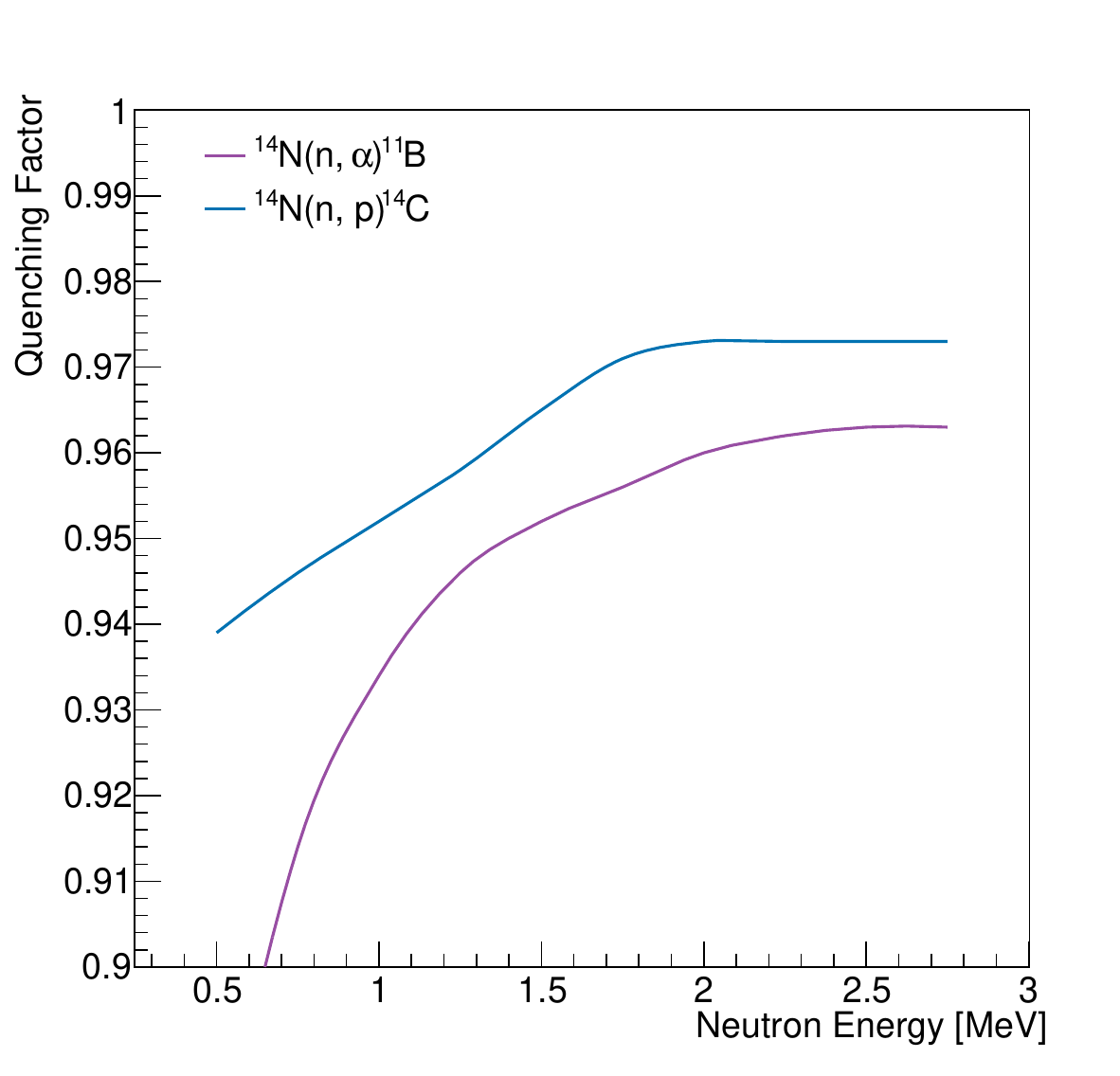}
    \caption{Ionisation quenching factor for each of the reaction channels as a function of neutron energy.
    \label{fig:qf}}
\end{figure}

The neutron energy was reconstructed on an event-by-event basis from the measured pulse amplitudes.  
The reaction $Q$-values of $\SI{-159}{\kilo\electronvolt}$ and $+\SI{625}{\kilo\electronvolt}$, for the \ce{^{14}N(n,\alpha)^{11}B} and \ce{^{14}N(n,p)^{14}C} reactions respectively, are taken into account in the reconstruction. The ionisation quenching factors are estimated using \textsc{SRIM}~\cite{ZIEGLER20101818},  accounting for energy distribution among reaction products. The overall ionisation quenching factors used are shown in Fig.~\ref{fig:qf}, and no further uncertainty is assigned. 

For both the \ce{^{14}N(n,\alpha)^{11}B} and \ce{^{14}N(n,p)^{14}C} channels, selection criteria in pulse rise time and amplitude are applied to suppress contributions from elastic neutron scattering, electronic noise, and the \ce{^{241}Am} source. Subsequently, the mean reconstructed energy  was determined from the energy distribution of the selected events. The uncertainty on the extracted mean was taken from the uncertainty returned by the Gaussian fit. 
The typical amplitude resolution was from \SI{3.1}{\percent} to \SI{5.0}{\percent} and $\sim$\SI{7}{\percent}  for the \ce{^{14}N(n,\alpha)^{11}B} and \ce{^{14}N(n,p)^{14}C}, respectively.
The systematic uncertainty associated with the event selection was evaluated by repeating the analysis with varied selection criteria and taking the maximum absolute variation of the extracted mean amplitude with respect to the nominal value. For the \ce{^{14}N(n,\alpha)^{11}B} channel, the minimum and maximum amplitude selections were varied. For the \ce{^{14}N(n,p)^{14}C} channel, both the minimum and maximum amplitude cuts and the minimum and maximum rise time selections were varied. For the \ce{^{14}N(n,\alpha)^{11}B} channel, the statistical and systematic uncertainties are of the same order, both being approximately \SI{1}{\percent}. For the \ce{^{14}N(n,p)^{14}C} reaction, the statistical uncertainty reaches a maximum of approximately \SI{1}{\percent}, while the systematic uncertainty reaches a maximum of \SI{8.5}{\percent}. In both cases, these maximum uncertainties are observed for the run collected at \SI{1.800}{\mega\electronvolt}.

\begin{figure}[htbp]
\centering
\includegraphics[width=0.75\linewidth]{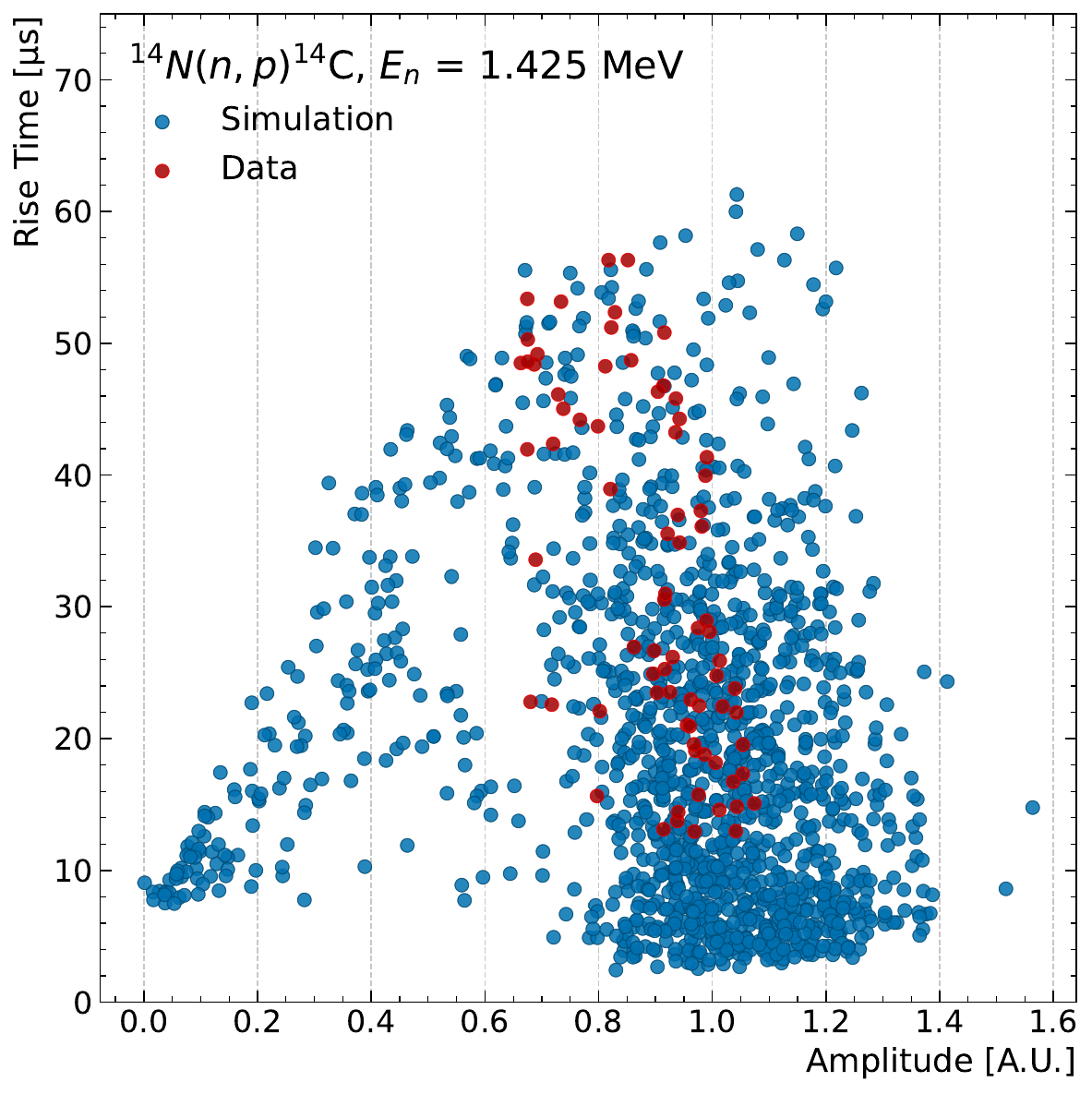}
\caption{Rise time versus amplitude for \ce{^{14}N(n,p)^{14}C} events.   
\label{fig:ballisticDef}}
\end{figure}

The ballistic-deficit correction for the \ce{^{14}N(n,p)^{14}C} channel was derived from an LTspice simulation \cite{ltspice2026}, in which the simulated current signals generated by the reaction were used as input to the electronics model. The resulting output amplitude was evaluated as a function of the signal rise time, as shown in Fig.~\ref{fig:ballisticDef}. The dependence of the reconstructed amplitude on the rise time observed in the simulation is consistent with the trend observed in the experimental data for the events tagged as containing a proton. The correction derived from the simulation was applied to the measured amplitudes, and the corrected mean amplitude was then extracted from the data. 
Since it is not established that the simulated correction should be applied to all events in the sample, the difference between the mean amplitude obtained without applying the correction and that obtained after applying it was assigned as an additional systematic uncertainty due to the ballistic deficit. Therefore, for the event identified as \ce{^{14}N(n,p)^{14}C} reaction, the systematic uncertainty increases to approximately \SI{10}{\percent} at all neutron energies.

\begin{figure}[htbp]
\centering
\includegraphics[width=0.85\linewidth]{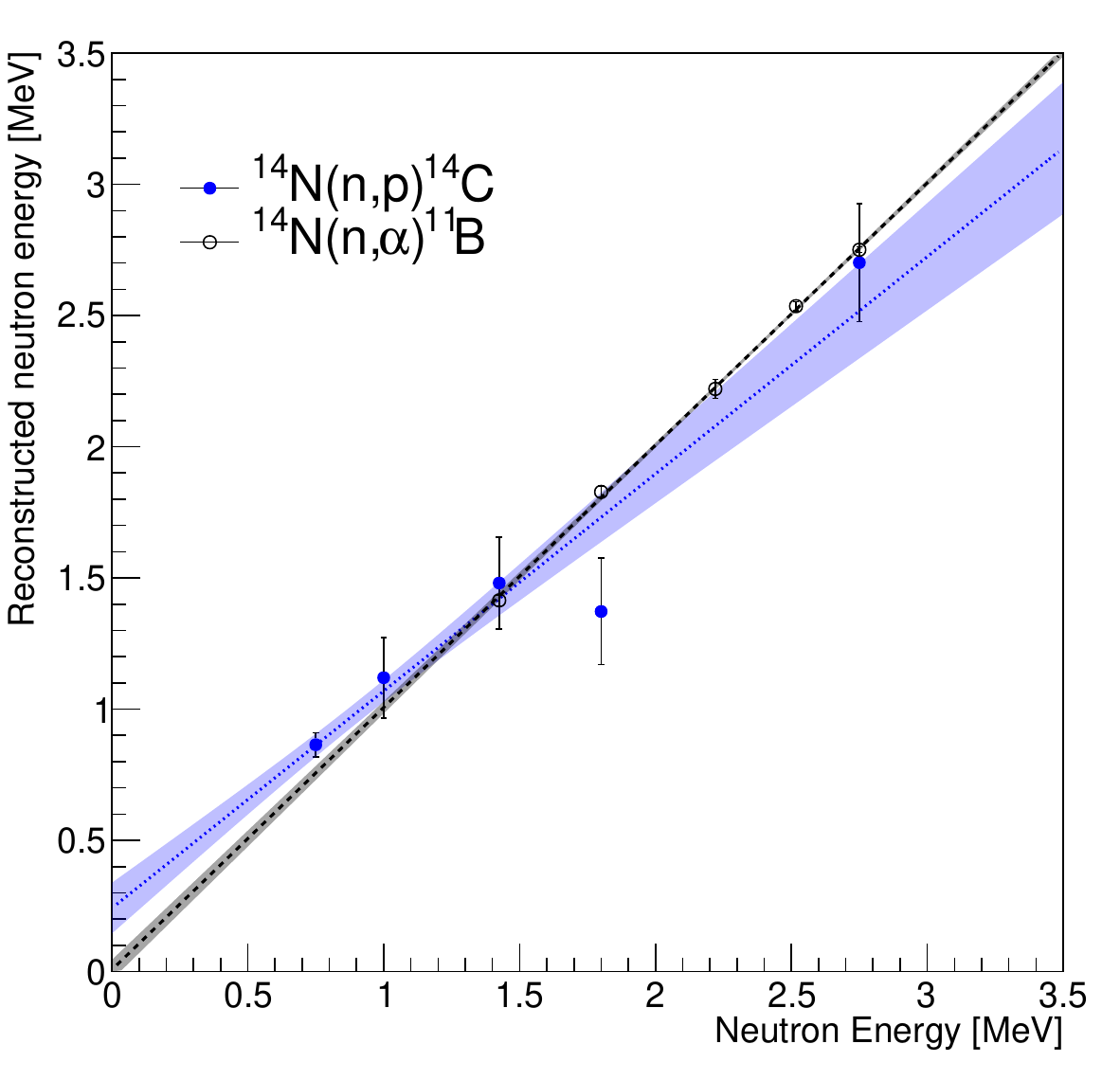}
\caption{Reconstructed neutron energy versus neutron beam energy for the \ce{^{14}N(n,p)^{14}C} and \ce{^{14}N(n,\alpha)^{11}B} reactions. \label{fig:recoEn}}
\end{figure}

Figure \ref{fig:recoEn} shows the measured neutron energy from \ce{^{14}N(n,p)^{14}C} and \ce{^{14}N(n,\alpha)^{11}B} reactions as a function of the neutron beam energy measured by the beam facility.  
The two distributions are fitted with a linear function and the obtained parameters are reported in Table \ref{tab:fit-results},
\begin{table}[!ht]
\centering
\caption{Reconstructed energy linearity.\label{tab:fit-results}}
\resizebox{\linewidth}{!}{%
\begin{tabular}{cccc}
\hline
Reaction & $E_0$ & $\frac{d E_{reco}}{d E_{beam}}$ & $\chi ^2 /\rm{NDF}$ \\
\hline
\hline
\ce{^{14}N(n,p)^{14}C} & $0.24\pm 0.10$ & $0.83 \pm 0.10$ & $4.04/3$ \\
\ce{^{14}N(n,\alpha)^{11}B} & $0.01\pm 0.04$ & $1.000 \pm 0.015$ & $2.47/3$ \\
\hline
\end{tabular}%
}
\end{table}
demonstrating the energy linearity of the spherical proportional counter in the energy range probed.

\section{Conclusions}
\label{sec:conclusions}
 
The first detection and spectroscopic measurement of mono-energetic neutrons with a nitrogen-filled spherical proportional counter at an accelerator facility has been presented. The measurements were performed at the \SI{5.5}{\mega\volt} Tandem accelerator of the National Centre for Scientific Research ``Demokritos'', using mono-energetic neutron beams of $E_n =$ \SIrange{0.75}{2.75}{\mega\electronvolt} produced via the \ce{^7Li(p,n)^7Be} reaction. 
The detector used was a 30 cm in diameter aluminium vessel filled with nitrogen and equipped with an 11-anode ACHINOS multi-anode sensor with individual anode read-out.
Neutrons were observed at each of the seven energies, and the incident neutron energy was reconstructed. The linearity of the detector response was established within the statistical uncertainties of the measurement.

In the longer term, this development opens the way to the deployment of the nitrogen-filled spherical proportional counter as a safe, cost-effective and reliable alternative to \ce{^3He}-based detectors in a variety of scientific, industrial and medical settings, including neutron background characterisation in underground laboratories hosting rare-event searches.

\section{Acknowledgments}
This work has been supported by the Deutsche
Forschungsgemeinschaft (DFG, German Research Foundation) under Germany's Excellence Strategy — EXC 2121 “Quantum Universe” — 390833306, and by the UKRI’s Science \& Technology Facilities Council through the University of Birmingham
Particle Physics consolidated grant (ST/W000652/1, UKRI/ST/C002848/1), and LM's PhD scholarship (ST/X508913/1). 
The authors thank Dr. Vasileios Vlachakis for performing the topographic survey of the experimental arrangement. 
The authors are grateful to the technical and operations staff of the Tandem Accelerator Laboratory at NCSR ``Demokritos'' for their support during the measurements.

\bibliographystyle{elsarticle-num}

\end{document}